 \documentclass[prd,twocolumn,letterpaper]{revtex4}
 \usepackage{graphicx}
 \usepackage{epstopdf}
 \usepackage{amsmath,amssymb}
 \usepackage{bm}
\newcommand{\apjl}{ApJ}

\newcommand{\aj}{AJ}
\newcommand{\mnras}{MNRAS}
\newcommand{\physrep}{Physics Reports}

\newcommand{\lya}{Ly$\alpha$ }
  \newcommand{\iMpc}{\mbox{ Mpc$^{-1}$}}
  
   \newcommand{\eV}{\mbox{ eV}}
 \newcommand{\yhe}{$Y_{\rm He}$ }
\begin{document}
 
   \title{Constraining massive neutrinos using cosmological 21 cm observations}
 
 \author{Jonathan R.~Pritchard}\thanks{Hubble Fellow}
 \email{jpritchard@cfa.harvard.edu}
\affiliation{Harvard-Smithsonian Center for Astrophysics, MS-51, 60 Garden St, Cambridge, MA 02138, USA}

\author{Elena Pierpaoli}
\affiliation{Physics and Astronomy Department, University of Southern California, Los Angeles, California 90089-0484, USA}

 %------------------------------------------------------------------------------
 
 %%%%%%%%%%%%%%%%%%%%%%%%%%%%%%%%%%%%%%%%%%%%%%%%%%%%%%%%%%%%%%%%%%%%%%%%%%%%%%%%
 \begin{abstract}

Observations of neutrino oscillations show that neutrinos have mass.  However, the best constraints on this mass currently come from cosmology, via measurements of the cosmic microwave background and large scale structure.  In this paper, we explore the prospects for using low-frequency radio observations of the redshifted 21 cm signal from the epoch of reionization to further constrain neutrino masses.  We use the Fisher matrix formalism to compare future galaxy surveys and 21 cm experiments.  We show that by pushing to smaller scales and probing a considerably larger volume 21 cm experiments can provide  stronger constraints on neutrino masses than even very large galaxy surveys.  Finally, we consider the possibility of going beyond measurements of the total neutrino mass to constraining the mass hierarchies.  For a futuristic, 21 cm experiment we show that individual neutrino masses could be measured separately from the total neutrino mass.

\end{abstract}
 
 % \keywords{cosmology}
 %------------------------------------------------------------------------------
 % User-supplied List of keywords.
 
% \pacs{98.80.k,98.80.Es}
 
\maketitle

  %%%%%%%%%%%%%%%%%%%%%%%%%%%%%%%%%%%%%%%%%%%%%%%%%%%%%%%%%%%%%%%%%%%%%%%%%%%%%%%%
\section{Introduction} 
\label{sec:intro}

%Review of the field \citep{lesgourgues2006}
The potential for using cosmological perturbations to constrain particle physics in general %\citep{}
and neutrinos in particular
% \citep{} 
was pointed out more than twenty years ago \citep{Bond83}.
Recent cosmological data have indeed allowed much progress in constraining 
dark matter properties  (e.g. \cite{Pierpaoli98, Hannestad04}) and neutrino physics  
(\cite{Bell06,Goobar06,Pierpaoli03,Seljak05,Elgaroy03,Crotty03,Crotty04,Pierpaoli04PRL,Pierpaoli04}).

Perhaps the most interesting neutrino property that can be investigated with cosmology is its mass.
While solar and atmospheric neutrino oscillation experiments set a milestone for particle physics models by showing that neutrinos do have a mass, by measuring mass differences 
  ($\Delta m_{21}^2 \simeq 8 \times 10^{-5} {\rm eV}^2$ and  $\Delta m_{31}^2 \simeq 2 \times 10^{-3} {\rm eV}^2$;  \cite{Maltoni04})
they don't carry any information on the absolute neutrino mass scale.
Interesting limits on the neutrino mass scale may  come from  tritium beta decay in the near future (e.g. KATRIN), which will probe electron neutrino masses above $~0.2\ {\rm eV}$.

Recent cosmological data, however, also provide the opportunity to probe neutrino masses through the
peculiar behavior of massive neutrinos, which are relativistic for part of the Universe's history and possibly not at later times. This behavior has an  impact on  both the background evolution and the growth of structures on intermediate and small scales.
The constraints are typically derived by combining information   on the amplitude of cosmological perturbations at the redshift of recombination (probed by the Cosmic Microwave Background) and in the very local Universe (probed by galaxy surveys such as the 2dF or SDSS).
Cosmology currently sets the tightest upper  limit on the neutrino mass (e.g. \cite{Seljak06,Goobar06}) to be about  $0.62\ {\rm eV}$. 
 It should be noted here  that this same mass limit  was recently derived with  the WMAP--5 year data 
combined with limits from supernovae (SN) and baryon acoustic oscillations (BAO) \cite{komatsu2008}.
With future surveys, alternative techniques will be available to determine neutrino mass, including the use of CMB--LSS cross correlation \cite{Lesg07} and lensing studies 
\citep{lesgo06lensing}.

While most methods mentioned above  make use of the fluctuation amplitude either at high redshifts ($z \simeq 1000$) or at low ones ($z \simeq 1$), a new cosmological observable has emerged in the past few years as a potential probe of fluctuations at intermediate redshifts ($ 6 \le z \le 20$): the study of fluctuations in the  21 cm line. 
As the power spectrum of perturbation at these redshifts is affected by the presence of massive neutrinos, the study of the 21 cm line may provide complementary information on neutrino masses \citep{McQuinn06}. Moreover, as fluctuations are smaller at higher redshifts, the effects of non--linearity 
at small scales are weaker than at the present time, allowing for a more straightforward interpretation of the data and ultimately to the use of a broader scale range.

As the Universe is still matter dominated at such high redshifts, cosmological probes of neutrino masses with the study of the 21 cm line  also offer the advantage of avoiding the typical degeneracy 
between  neutrino mass and  dark energy which is seen in probes relying on power spectrum measurements at redshifts $z \le 1$ \citep{Hannestad05}.
The study of 21 cm fluctuation  would, however, still exploit 
 the impact of neutrinos on the matter power spectrum by probing perturbation amplitudes at redshifts  substantially different from the ones probed by CMB studies.

The cosmological limits cited above were derived under the hypothesis that neutrinos all share the same mass. 
As neutrino oscillations only determine two of the three mass square differences, we do not know the 
precise hierarchy for the masses ($m_1<m_2<m_3$  or $m_3<m_1<m_2$ representing normal and inverted hierarchy respectively). The relation between the mass of the lightest neutrino and the sum of all neutrino masses clearly depends on the particular hierarchy, however this relation is affected 
by the hierarchy only if the lightest neutrino has a mass below $\simeq 0.3 \rm{eV}$. 
Also, for the time being, the effect of different neutrino masses has been largely irrelevant for cosmological considerations as the upper limits on the mass imposed by cosmology are well above the mass splitting.
As cosmological data  become more and more abundant and different probes contribute in determining the neutrino mass, it is possible that cosmology will be sensitive to the masses, which are comparable to the measured mass splitting. In such a case, cosmology would, not only, set the tightest measurement of the neutrino mass scale, but could also enable one to distinguish between the normal and inverted hierarchy (hereafter NH and IH respectively).

In this paper, we consider the potential of constraining neutrino masses with future surveys mapping the 21 cm line. We will investigate how these measurements complement and improve the results 
based on the future CMB probes (Planck) and large scale structure surveys.
In particular, as previous studies on this topic suggest that the precision in the determination of the neutrino mass should be lower than the mass splitting \cite{Mao08}, we study the potential of future 21 cm surveys to 
determine the neutrino hierarchy.

This paper is organized as follows: in sec.~\ref{sec:21cmgen} we discuss the generalities of the 21 cm probe; in sec.~\ref{sec:nugen} we discuss how the neutrino mass specifications affect the power spectrum at the relevant scales and redshifts, in sec.~\ref{sec:fisher} we present the adopted Fisher matrix approach, while sec.~\ref{sec:results} and \ref{sec:conc} are dedicated to results and conclusions respectively.

%%%%%%%%%%%%%%%%%%%%%%%%%%%%%%%%%%%%%%%%%%%%%%%%%%
%%%%%%%%%%%%%%%%%%%%%%%%%%%%%%%%%%%%%%%%%%%%%%%%%%

\section{21 cm as a cosmological probe}
\label{sec:21cmgen}

We begin by briefly summarizing the physics of the 21 cm signal and refer
the interested reader to \citet*{fob} for further details.  The 21 cm line
of the hydrogen atom results from hyperfine splitting of the $1S$ ground
state due to the interaction of the magnetic moments of the proton and the
electron.  The HI spin temperature $T_S$ is defined through the ratio
between the number densities of hydrogen atoms in the $1S$ triplet and $1S$
singlet levels, $n_1/n_0=(g_1/g_0)\exp(-T_\star/T_S)$, where $(g_1/g_0)=3$
is the ratio of the spin degeneracy factors of the two levels, and
$T_\star\equiv hc/k\lambda_{21 \rm{cm}}=0.068\,\rm{K}$.  
The optical depth of this transition is small at all relevant redshifts,
yielding brightness temperature fluctuations
%\begin{equation}\label{brightnessT}
\begin{multline}\label{brightnessT}
T_b=27  x_{\rm{HI}}\left(1+\frac{4}{3}\delta_b\right)\left(\frac{\Omega_bh^2}{0.023}\right)\left(\frac{0.15}{\Omega_mh^2}\frac{1+z}{10}\right)^{1/2}\\ \times\left(\frac{T_S-T_\gamma}{T_S}\right)\,\rm{mK}.
\end{multline}
%\end{equation}
Here $x_{\rm{HI}}$ is the neutral fraction of hydrogen, $\delta_b$ is the
fractional overdensity in baryons, and the factor of $4/3$ arises from
including the effect of peculiar velocities and averaging the signal over
angles.  Provided that there is a sufficient background of \lya photons, the spin temperature $T_S$ is coupled to the gas kinetic temperature $T_K$ via the Wouthysen-Field effect \citep{wouth1952,field1958}.

In general, $T_b$ is sensitive to details of the IGM and is complex to model.  However, with a few reasonable assumptions this dependence drops out.  At redshifts long after star formation begins, it is likely that the X-ray background produced by early stellar remnants has heated the inter-galactic medium (IGM) so that the gas kinetic temperature $T_k$ is much greater than the CMB temperature $T_{\rm \gamma}$ \citep{mmr1997,furlanetto2006,pritchard2008}.  The same star formation produces a large background of \lya photons sufficient to couple $T_S$ to $T_K$.  In this scenario, appropriate for $z\lesssim10$, we are justified in taking $T_S\approx T_K\gg T_{\gamma}$ in which limit $T_b$ shows no dependence on $T_S$.  We need then only worry about fluctuations in the density and neutral fraction. If the IGM is fully neutral, 21 cm fluctuations arise only from the variation in density from point to point and $P_{21}$ provides a tracer of the matter power spectrum.   In this limit, we can write the power spectrum of 21 cm brightness fluctuations as
\begin{equation}
P_{21}(k)=T_b^2P_{\delta\delta}(k)(1+\mu^2)^2,
\end{equation}
where $P_{\delta\delta}$ is the matter power spectrum and $\mu$ is the angle between the line of sight and the wavenumber $\mathbf{k}$.
In this situation, 21 cm observations may be thought of as being analogous to galaxy surveys with $T_b$ replacing the unknown galaxy bias.  While in principle $T_b$ can be calculated given a knowledge of cosmological parameters, its dependence on the largely unknown ionization and thermal history means that we should not treat it as a known quantity, but instead as a free parameter to be measured.  Thus cosmological information will come primarily from the shape of $P_{21}$ and not its amplitude.

Throughout this paper, we will adopt the above optimistic assumption that 21 cm observations at $z\lesssim10$ measure the underlying matter power spectrum modulated by the 21 cm brightness temperature, i.e. $P_{21}(k)\propto T_b^2P(k)$.  In practice, this is unlikely to be the case since a combination of fluctuations in the Ly$\alpha$ flux \citep{bl2005detect,pritchard2006}, inhomogeneous X-ray heating \citep{pritchard2007xray}, and inhomogeneous reionization \citep{fzh2004} will modify the 21 cm power spectrum significantly.  Although these fluctuations are unlikely to all contribute at the same time \citep{pritchard2008}, finding a redshift range where only the density fluctuations are significant would be extremely fortuitous.  In principle, the matter power spectrum may be measured in the presence of these other contributions by precise measurements of the angular dependence of the 21 cm power spectrum \citep{bl2005sep}.  This
 procedure, however, reduces the information extracted, therefore
 the constraints on cosmological parameters that may be achieved in this pessimistic scenario are seriously weakened \citep{mao2008}.  More likely detailed modeling of reionization will allow cosmological parameters to be extracted in the presence of neutral fraction fluctuations with only moderate degradation $\sim 10-50\%$ in the constraints \citep{mao2008}.  Finally, since the majority of these extra sources of fluctuation are most significant on large scales, on small scales $k\gtrsim1\iMpc$ it might be hoped $P_{21}(k)$ will trace $P(k)$ closely (although see \citet{lidz2007ho}).

We note that, although the details of reionization are still poorly constrained, increasingly accurate observations of the CMB constrain the optical depth to the surface of last scattering.  This provides an integral constraint on the ionization history that can be used to limit when reionization ends.  The most recent WMAP results \citep{komatsu2008} give $\tau=0.09\pm0.017$.  If reionization is assumed to end instantaneously this translates into $z_{\rm reion}=10.8\pm1.4$, while more extended reionization models lead to reionization completing at lower redshifts (e.g. using the model of \citet{furlanetto2006} gives $z_{\rm reion}=9\pm1.4$).  A complementary constraint on the end of reionization comes from observations of the Gunn-Peterson trough in the \lya forest in absorption spectra of quasars at redshift $z\gtrsim6.5$ \citep{becker2001,fan2006}.  This is widely interpreted as indicating a neutral fraction of $x_{HI}\sim10^{-5}$ in the IGM, possibly signalling the tail end of reionization.  Guided by these observations, we will model reionization as occurring sharply at $z=7.5$.  Although simplistic, this sets a reasonable lower limit on 21 cm observations during the epoch of reionization.  Although some neutral hydrogen remains in dense regions after reionization and may be targeted by future 21 cm observations \citep{wyithe2007}, the post-reionization signal strength decreases significantly.  For this reason, we will restrict our 21 cm experiments observations to $z\gtrsim7.5$.  The upper limit on the accessible redshift range comes from the experimental design.  For currently proposed experiments, this is likely to be around 100 MHz ($z=13$).  Given the rapidly increasing foregrounds and uncertainty in the spin temperature at higher redshifts, we will limit ourselves to redshifts $z\lesssim 10$.  In principle, increasing the redshift range will increase parameter constraints, but the increasing thermal noise reduces the utility of additional high redshift bins.  

%%%%%%%%%%%%%%%%%%%%%%%%%%%%%%%%%%%%%%%%%%%%%%%%%%
%%%%%%%%%%%%%%%%%%%%%%%%%%%%%%%%%%%%%%%%%%%%%%%%%%

\section{Effect on neutrino masses on the power spectrum}
\label{sec:nugen}
Although the physics of neutrinos is potentially very rich and models exist for which the number of neutrino species is different from three, 
for the purpose of this paper, we will only consider the standard case with  three light neutrino species, and will discuss 
the implications of having different assortments of neutrino masses.

\begin{figure}
\centering
\includegraphics[width=8.6cm,height=8.6cm]{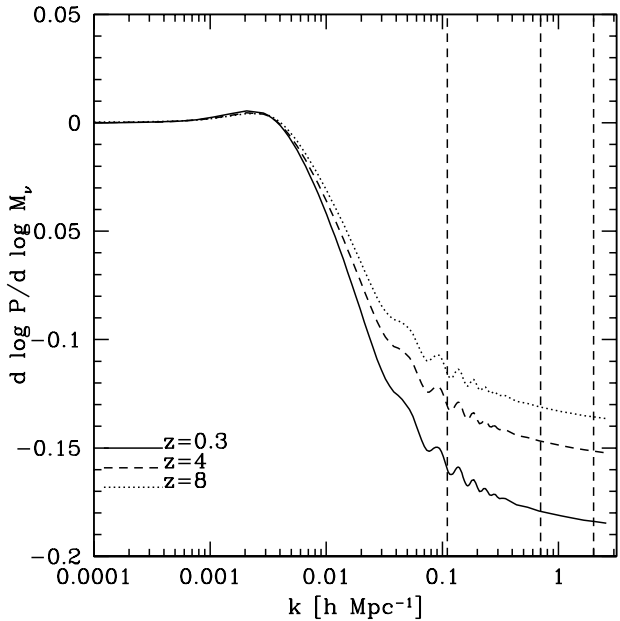}
\caption{\label{fig:dpk}  Effect of  a  different neutrino mass on the power spectrum at redshifts $z=0.3$ (solid curve), 4 (dashed curve), and 8 (dotted curve).  The curves represent the derivatives of the matter power spectrum with respect to neutrino mass for a fiducial mass value of $M_\nu=0.3\eV$. The vertical dashed lines represent the non-linear scale at redshifts $z=0.3$, 4, and 8 (from left to right).  Note that the derivative becomes more negative at lower redshifts, as massive neutrinos have more time to affect the power spectrum.}
\end{figure}

If neutrinos were in thermal contact with the CMB at early times, their number density is fixed.
Therefore, their  contribution to the  total present energy density is strictly related to their masses:
$ \Omega_\nu = \sum_i m_i/ 93.14 h^2 {\rm eV}$ and they would contribute a fraction $f_\nu = \Omega_\nu / \Omega_m$ to the total mass density of the Universe. 
From neutrino oscillations, we know that at least two neutrinos are non--relativistic today, as 
the mass differences implied are about 0.047 and 0.009 eV. 
The current limit on the neutrino mass scale  is about  $0.6\ {\rm eV}$.  If neutrinos have masses above $1.7 \times 10^{-4} {\rm eV}$ (that is the neutrino temperature today) and below the current mass scale limit, they are still relativistic when cosmologically relevant scales enter the horizon.
Therefore, because of their free--streaming behavior, they then tend to erase fluctuations on such scales. 
For a given neutrino mass, the impact of the free--streaming behavior on the actual matter power spectrum depends on the scale considered:  scales larger than the horizon size when neutrinos become non-relativistic 
are unaffected; while smaller scales suffer a damping. Assuming that neutrinos contribute to the expansion of the Universe, but not to the clustering,
the growth of cold dark matter density perturbations as a function of time  is: $\delta_{CDM} \propto a^{1-(3/5)f_\nu}$. For small 
values of the neutrino mass, this effect is tiny; However, it may last for a very extended period of time, resulting in a sizable decrement of the matter power spectrum.
Since the CMB probes a perturbation's amplitude at decoupling ($z \simeq 1000$) and the 21 cm measurements
probe much lower redshifts ($z \simeq 6-10$) the effect  can become  significant.
%The  suppression  of the power spectrum at the present time  is approximated by the familiar formula: $P(k)^{f_\nu}/P(k)^{f_\nu = 0} = -8 f_\nu$. 
Free--streaming is not the only physical effect caused by massive neutrinos impacting the matter  power spectrum. As neutrinos are relativistic at early times and become non--relativistic at late times, 
they  contribute to the radiation energy density initially and to the matter energy density at  later times.
Therefore, according to the actual values of their masses, they may slightly change the redshift  of matter-radiation equivalence; which in turn has an impact on when dark matter perturbations start growing at the rate of  a matter 
dominated Universe.
Fig. \ref{fig:dpk} shows the effect of neutrino mass on the  power spectrum at different scales and redshifts,  by looking at the derivative of the power spectrum with respect to $M_\nu$ assuming the three species have total mass $M_\nu=0.3\eV$.
Smaller scales are the most affected by different neutrino masses. Such scales tend to be non--linear at the present time,
%EP
but still linear at $z=8$.
The 21 cm probe can therefore take advantage from the information provided by all such small scales.

\begin{figure}
\centering
\includegraphics[width=8.6cm,height=8.6cm]{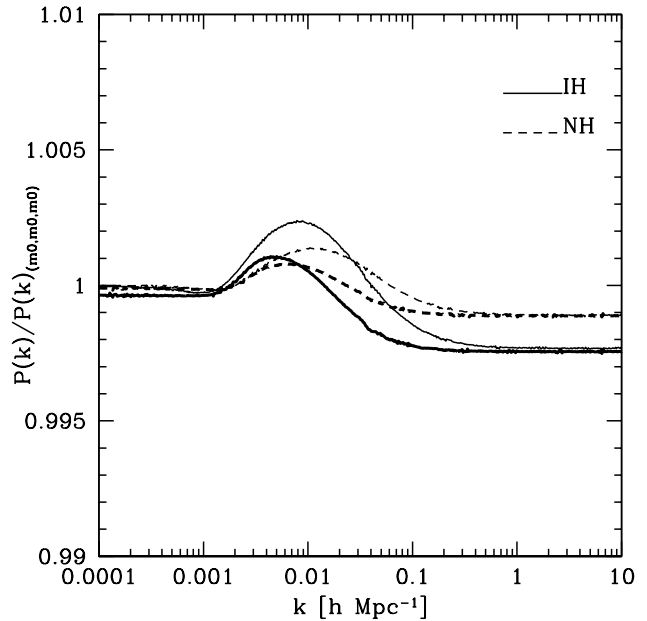}
\caption{\label{fig:ratio}  Difference between the power spectra between the normal and inverted hierarchy at different redshifts. We plot the ratio of the power spectrum calculated with total neutrino mass $M_\nu=0.12$ distributed in the normal hierarchy (dotted curve) and inverted hierarchy (solid curve) relative to the case where the three neutrino masses are degenerate ($m_1=m_2=m_3$).  These curves are calculated at $z=0$ (thin curves) and $z=8$ (thick curves).  The difference continues to smaller scales, but saturates at $\sim0.2 \%$ at $ k=1  h {\rm Mpc}^{-1}$.}
\end{figure}

Apart from the general effects implied by neutrinos having a mass, should the overall mass scale be 
below $ \simeq 0.3\  \rm{eV}$, then the behavior of each single neutrino species may impact the power spectrum.
This hypothesis has been investigated in detail in \cite{lesgourgues2004}, which  shows maximum differences of the order of $0.5\%$ in the present power spectrum range $0.001 \le k \le 0.1~ h {\rm Mpc}^{-1}$ for different mass configurations and a total  fiducial mass of 0.12 eV. Such small differences  cannot be probed with 
current galaxy surveys like SDSS, however they may be with future  cosmic shear surveys.
We argue here that future 21 cm surveys may well also allow us to distinguish between different mass configurations, in particular, it may enable us to distinguish between the normal and inverted hierarchy.
Fig.~\ref{fig:ratio} represents the difference in the power spectra between the normal and inverted hierarchy relative to the case of three neutrinos with degenerate masses at redshifts $z=0$ and $z=8$, of interest for the 21 cm probe. 
Although differences in the power spectrum are tiny ($0.1-0.2 \%$), at $z=8$ they extend in the precise $k$ range probed by the 21 cm surveys: $ 0.1 \le k \le 2 ~h {\rm Mpc}^{-1}$.
In the following, we will discuss which kind of surveys may reach the sensitivity to neutrino masses 
which allow the determination of the neutrino mass hierarchy.

Another advantage of studying the matter perturbation evolution through the 21 cm  line is that  at the relevant times 
the Universe is still matter dominated. At lower redshifts $z \le 2$, dark energy dominates the expansion and influence the perturbations' growth, slowing it down with respect to the matter dominated regime. The actual growth rate depend 
on the specific value of $w(z)$, the equation of state parameter for dark energy, which is one of the least currently constrained parameters.  As a result, 
for a given perturbation' s amplitude  at the time of recombination, an  observed value at the present time  showing a slower  evolution  than that implied by a matter dominated model may  be interpreted as either a (small but long--lasting) neutrino  effect or a (big, late time) dark energy one.
There is currently no way of decoupling 
the damping effect of neutrino  free--streaming on the power spectrum from the dark energy effect,
unless other probes not involving growth of perturbations can characterize the evolution of dark energy 
very well.

We note here that current limits on neutrino masses at the level on $M_\nu \simeq 0.6$ eV are not only reached using galaxy surveys and the Lyman--$\alpha$ forest, but also using WMAP-5 yrs in combination with BAO and SN measurements,  therefore completely avoiding the use of the matter power spectrum \cite{komatsu2008}. In this case, the degeneracy between the neutrino mass determination and  the quintessence equation of state no longer exists.  Additionally, 21 cm studies will be subject to a completely different set of systematic uncertainties to galaxy surveys and the Lyman-$\alpha$ forest.  Exploiting multiple techniques to probe the matter power spectrum on small scales will likely be necessary to ensure that the various systematics of different techniques are under control.

However,  it is unlikely that the CMB alone will ever greatly improve on this constraint; leaving the fine measurements of neutrino masses to some kind of probe involving also  the matter power spectrum. The 21 cm probe capitalizes on the effect of neutrino masses on small scale perturbations while avoiding the low--redshift complications implied by the  dominance of  the currently poorly--characterized dark energy and also the greater non-linearities existing on small scales.

Let us finally comment on the role of the CMB in determining the neutrino mass. Given that the CMB
decouples when the temperature of the Universe was 0.26 eV, it cannot be sensitive to very small neutrino masses, with the possible exception  of when the lensing effect by lower redshift structures is taken in consideration.  However, the upcoming Planck experiment has an incredible constraining power on generally all cosmological parameters, which is hardy attained by any other single cosmological probe. 
In turn, Planck will enable other probes (like galaxy surveys and the 21 cm) to attain a higher precision 
in the neutrino mass determination by providing accurate measurements for other cosmological parameters like $\Omega_m$ and $H_0$ with which the neutrino mass may be partially degenerate.
We will therefore use Planck as a baseline and will consider its projected performances 
in our calculations.

  %%%%%%%%%%%%%%%%%%%%%%%%%%%%%%%%%%%%%%%%%%%%%%%%%%%%%%%%%%%%%%%%%%%%%%%%%%%%%%%%
\section{Fisher calculation} 
\label{sec:fisher}

To make predictions for parameter constraints from future surveys we make use of the Fisher matrix formalism.  This has been described in detail elsewhere \citep{TTH1997} and we provide only a brief summary here of the details relevant for our analysis.

We take as our basic set of parameters the three parameters describing the energy density in matter $\Omega_mh^2$, baryons $\Omega_bh^2$, and dark energy $\Omega_\Lambda$, two parameters describing the tilt $n_s$  and amplitude of the primordial inflationary power spectrum $A_S$, and the optical depth $\tau$ to the surface of last scattering.  In addition to these basic parameters, we add the total mass in neutrinos $M_\nu=m_1+m_2+m_3$ and, since it modifies power on small scales the helium fraction $Y_{He}$.  We assume a flat Universe throughout.  This gives a parameter vector
$\Theta=(\Omega_m h^2, \Omega_b h^2, \Omega_\Lambda, n_s, A_s^2, \tau, Y_{\rm He}, M_\nu)$.  In addition, we require an effective bias -- either a genuine bias for galaxy surveys or the brightness temperature for 21 cm experiments.  We will separately explore the degeneracies introduced by allowing the equation of state $w$ of the dark energy to vary.  Throughout, we use CAMB \citep{Lewis:1999bs} to calculate CMB anisotropies and matter power spectra.  Our fiducial set of parameters is $\Theta=(\Omega_m h^2, \Omega_b h^2, \Omega_\Lambda, n_s, A_s^2, \tau, Y_{\rm He}, M_\nu)=(0.147,0.023,0.7,0.95,26,6,0.1,0.24,0.3\,\eV)$, which yields $\sigma_8=0.77$, consistent with the latest WMAP data \citep{komatsu2008}.

To evaluate $F_{ij}$, we need to specify a model, which determines the dependence of the likelihood function on $\Theta$, and a point in parameter space where we wish to determine parameter uncertainties.   In the case that the model parameters are Gaussian distributed, the Fisher matrix takes the form 
  \begin{equation}\label{fisher_general}
F_{\alpha\beta}=\frac{1}{2}\rm{\rm{tr}}(C^{-1}C,_\alpha C^{-1}C,_\beta)+\frac{\partial\mu}{\partial\theta_\alpha}C^{-1}\frac{\partial\mu}{\partial\theta_\beta},
\end{equation}
where $C$ is the covariance matrix for the data, and $\mu$ is the data's mean.  This will be a good approximation in the case of CMB observations, galaxy surveys (see \cite{Hannestad07} for a discussion on full likelihood constraints), and 21 cm observations.  Note that, for our purposes, we will need to combine information from these three sources.  When used together these data sets break many degeneracies that are present when they are used alone.  Let us consider the Fisher matrix from each in turn.

A CMB experiment may be characterized by a beam size $\theta_{\rm{beam}}$ and sensitivities to temperature $\sigma_T$ and polarization $\sigma_P$.  Given these quantities, the Fisher matrix is given by \citep{jungman1996,jungman1996L,kks,zaldarriaga1997}
 \begin{equation}\label{fisher_cmb}
F^{CMB}_{\alpha\beta}=\sum_\ell\sum_{X,Y} \frac{\partial C_\ell^X}{\partial\theta_\alpha} (\rm{Cov}_\ell)^{-1}_{XY}\frac{\partial C_\ell^Y}{\partial\theta_\alpha},
\end{equation}
where $C_\ell^X$ is the power in the $\ell$th multipole for $X=T,E,B,$ and $C$,  the temperature, E-mode polarization, B-mode polarization, and TE cross-correlation respectively.  The elements of the covariance matrix $\rm{Cov}_\ell$ between the various power spectra are given in \citep{kks,zaldarriaga1997}.  Since in the near future the best CMB observations are likely to come from the Planck satellite, we include specifications for that satellite in Table \ref{tab:cmb}.  We will also consider a cosmic variance limited experiment (CosmicVar) to indicate the ultimate precision obtainable from CMB observations.  In each case, we consider information from $l\le2500$, assuming that  scales contaminated by secondary anisotropies such as the Sunyaev-Zeldovich effect can be cleaned.  We also exclude any information from B-mode polarization, although the B-mode signal generated by weak gravitational lensing of E-mode polarization does in principle contain information on neutrinos via the lensing potential power spectrum.

 \begin{table}
\caption{Specification for CMB experiments.  For each channel we list the frequency in GHz, the beam size $\theta_{\rm{beam}}$ is the FWHM in arcmin.  Sensitivities $\sigma_T$ and $\sigma_P$ are in $\mu \rm{K}$ per FWHM beam, $w=(\theta_{\rm{beam}}\sigma)^{-2}$. Taken from \citet{lesgourgues2004}.  In each case, we assume $f_{\rm sky}=0.65$ allowing for subtraction of the galactic plane.}
\begin{center}
\begin{tabular}{c|c|c|c|c}
Experiment & Frequency& $\theta_{\rm{beam}}$ & $\sigma_T$ & $\sigma_P$ \\
\hline
Planck & 100 & 9.5 & 6.8 & 10.9 \\
         & 143 & 7.1 & 6.0 & 11.4 \\
	& 217 & 5.0 & 13.1 & 26.7 \\
\end{tabular}
\end{center}
\label{tab:cmb}
\end{table}%

Moving now to galaxy surveys, we may write the appropriate Fisher matrix as
\citep{tegmark1997,se2003probe}
\begin{equation}\label{fisher_gal}
F^{GAL}_{\alpha\beta}=\int_0^{k_{\rm{\rm{max}}}}\frac{\partial\ln P(\mathbf{k})}{\partial\theta_\alpha}\frac{\partial\ln P(\mathbf{k})}{\partial\theta_\beta}V_{\rm{eff}}(\mathbf{k}) \frac{d^3\mathbf{k}}{2(2\pi)^3},
\end{equation}
where the derivatives are evaluated using the cosmological parameters of the fiducial model and $V_{\rm{eff}}$ is the effective volume of the survey, given by
\begin{eqnarray}\label{veff}
V_{\rm{eff}}(k,\mu)&=&\int d^3\mathbf{r}\left[\frac{n(\mathbf{r})P(k,\mu)}{n(\mathbf{r})P(k,\mu)+1}\right]^2\nonumber\\
&\approx&\left[\frac{\bar{n}P(k,\mu)}{\bar{n}P(k,\mu)+1}\right]^2V_{\rm{survey}}.
\end{eqnarray}
Here the galaxy survey is parametrized by the survey volume $V_{\rm{survey}}$ and the galaxy density $n(\mathbf{r})$, which in the last equality we assume to be uniform $\bar{n}$ within a redshift bin.  We assign an amplitude $\sigma_{8,g}$ for clustering of galaxies on $8h\iMpc$ scales in each redshift bin.  This is used to calculate the galaxy bias $b=\sigma_{8,g}/\sigma_8$, where $\sigma_8$ is the clustering of the dark matter.  We take the observed power spectrum to be $P_{\rm obs}(k)=b^2 P(k)$, neglecting peculiar velocity effects.  Since the power spectrum is reconstructed from observations of galaxy positions using a fiducial cosmology $P(k)$ aquires additional cosmological dependence arising from the Alcock-Paczynski effect.  These help break some of the parameter degeneracies in the shape of the matter power spectrum by placing constraints on $H(z)$ and $D_A(z)$.  This is included in our analysis.

In addition, we must specify $k_{\rm{max}}$, a cutoff on small scales to avoid the effects of non-linearity.  We model this cut-off scale on the prescription of \citet{se2003probe}, who assumed that non-linear effects become important on a scale $k_{nl}=\pi/2/R_{nl}$ where $R_{nl}$ is the scale on which averaged density fluctuations $\sigma(R_{nl})=0.5$.  This represents a moderately conservative choice of cutoff.  The analysis of \citet{jeong2006} suggests that using 3rd order perturbation theory the matter power spectrum may be accurately modeled well into the quasi-linear regime significantly extending the accessible range of $k$-modes.  More recently, \citet{saito2008} have shown that, when the effect of neutrino masses on non-linear clustering is included in the calculation of the power spectrum by 1-loop perturbation theory, the power spectrum suppression in the non-linear regime is enhanced.  In this case, the combination of an enhanced signal and probing a larger volume by including the contribution of more scales increases the tightness of neutrino mass constraints by a factor of a few.  Including these effects may allow for improvement over our results and indicates some of the complications of recovering cosmological information from small scales.  Similar problems apply to the biasing of galaxies with respect to the background matter field, which may, on small scales, become scale dependent or be modified by the legacy of inhomogeneous reionization \citep{babich2005, pritchard2007bub}.  In this case, additional information from statistics beyond the power spectrum, e.g. the bispectrum \citep{fry1994,sefusatti2007}, may be necessary to quantify and correct for these effects.

Since much of the information about massive neutrinos shows up as small scale modifications to the power spectrum our choice of the cutoff $k_{\rm max}$ is important.  Too small a value and we will throw away important information about massive neutrinos, too large a value and we will include scales affected by non-linear evolution and will underestimate our uncertainty.  Note we will be considering 21 cm information, which comes from smaller scales still, but since this comes from higher redshifts the non-linear scale is smaller.

To indicate the spread of galaxy survey performance, we consider four galaxy surveys (with parameters modeled on those from \citep{se2003probe}).  Our basic galaxy survey, which we label SDSS, loosely corresponds to the SDSS LRG sample and is centered at $z\approx0.3$.  Next, we consider a larger galaxy survey centered at $z=1$, labeled G1.  With larger volume, higher galaxy density, and larger $k_{\rm max}$ this survey represents a considerable improvement over SDSS.  To illustrate the gain of going to higher redshift, we consider a Lyman break galaxy survey (G2) centered at $z=3$.  This has reduced volume, but compensates with higher galaxy density and larger $k_{\rm max}$.  Finally, we consider an all-sky galaxy survey out to $z=1$ (G3) capable of surveying $40 h^{-3}{\rm Gpc}^3$.  This represents the limit of galaxy surveys in the local Universe and is enormously better than the other surveys that we consider.  We list parameters for our galaxy surveys, along with our chosen cutoff scale, for each of our surveys in Table \ref{tab:galaxy}.   We will consider  surveys with spectroscopic redshifts, so that errors in galaxy redshifts are negligible.  Photometric surveys may achieve the same level of constraint, although they require much larger volumes.  The large redshifts errors greatly degrading sensitivity on small scales, which will hamper constraining neutrino masses.  For example, the Large Scale Synoptic Telescope (LSST) is a full sky photometric survey of volume exceeding that of our G3, but since it relies upon photometric redshifts provides parameter constraints only a factor of $\sim3$ better than G1.

\begin{table}
\caption{Specification for galaxy surveys. For each galaxy survey, we specify the mean redshift $z$, the survey volume $V_{\rm survey}$, the mean galaxy density $\bar{n}$ in the redshift bin,  the cutoff off scale $k_{\rm max}$, and the effective clustering of the galaxies $\sigma_{8,g}$.}
\begin{center}
\begin{tabular}{c|c|c|c|c|c}
 Survey& $z$ & $V_{\rm{survey}}$ & $\bar{n}$& $k_{\rm{\rm{max}}}$ & $\sigma_{8,g}$\\
 	&	&$(h^{-3} \rm{Gpc^3)}$&$(h^3\, \rm{Mpc}^{-3})$&$(h\, \rm{Mpc}^{-1})$&\\
\hline
SDSS & 0.3 & 1.0 &$10^{-4}$ & 0.1 &1.8\\
G1 & 1.0 & 1.8 &$5\times10^{-4}$ & 0.2 &1.0\\
G2 & 3.0 & 0.5 &$10^{-3}$ & 0.5 &1.0\\
G3& 1.0 & 40.0 &$5\times10^{-4}$ & 0.2 &1.0\\
\end{tabular}
\end{center}
\label{tab:galaxy}
\end{table}%

Estimates of the error on a 21 cm power spectrum measurement are calculated using the Fisher matrix \citep{mcquinn2005,bowman2006}
\begin{equation}\label{fisher_21cm}
F^{21CM}_{ij}=\sum_{\rm pixels} \frac{k^2 V_{\rm survey}}{4\pi^2}\frac{1}{\sigma_P^2(k,\mu)}\frac{\partial P_{T_b}}{\partial \theta_i}\frac{\partial P_{T_b}}{\partial \theta_j}.
\end{equation}
In this equation, $V_{\rm survey}=D^2\Delta D(\lambda^2/A_e)$ denotes the effective survey volume of our radio telescopes given the conformal
distance $D(z)$ to the center of the survey at redshift $z$, the depth of
the survey $\Delta D$, the observed wavelength $\lambda$, and the effective
collecting area of each antennae tile $A_e$.

The variance of a 21 cm power spectrum estimate for a single
$\mathbf{k}$-mode with line of sight component $k_{||}=\mu k$, restricting
ourselves to modes in the upper-half plane, including both sample variance
and thermal detector noise, and assuming Gaussian statistics, is given by
\citep{mcquinn2005,lidz2007}:
%\begin{equation}
\begin{multline}
\sigma_P^2(k,\mu)=\\ \frac{1}{N_{\rm field}}\left[\bar{T}_b^2P_{21}(k,\mu)+T_{\rm sys}^2\frac{1}{B t_{\rm int}}\frac{D^2\Delta D}{n(k_\perp)}\left(\frac{\lambda^2}{A_e}\right)^2\right]^2.
\end{multline}
%\end{equation}
The first term in this expression is the contribution from sample variance,
while the second describes the thermal noise of the radio telescope.  The
thermal noise depends upon the system temperature $T_{\rm sys}$, the survey
bandwidth $B$, the total observing time $t_{\rm int}$.  The effect of the
configuration of the antennae is encoded in the number density of baselines
$n_\perp(k)$ that observe a mode with transverse wavenumber $k_\perp$
\citep{mcquinn2005}.  Observing a number of fields $N_{\rm field}$ further reduces the variance.

In making predictions for 21 cm observations we will consider three separate instruments representing different levels of technological advancement.  As an experiment representative of a 21 cm pathfinder experiment, we consider the {\em Murtchison Widefield Array} (MWA), currently under construction in the Australian outback.  For this we take 500 antennae distributed with a filled core of size $20$m surrounded by the remainder of the antennae distributed with a $r^{-2}$ profile out to $750$m. To represent a mature 21 cm experiment, we consider an experiment modeled after the {\em Square Kilometer Array} (SKA).  For this, we distribute 20\% of a total of 5000 antennae within a 1km radius in the same way as the MWA -- a core surrounded by a $r^{-2}$ distribution.  These are then surrounded by a further 30\% of the total antennae in a uniform density annulus of outer radius 6km.  The remainder of the antennae are distributed at larger distances too sparsely to be useful for power spectrum measurements.  Finally, we consider a hypothetical {\em Fast Fourier Transform Telescope} \citep{fftt}, a future square kilometer collecting area array optimized for 21 cm observations.  In contrast to MWA and SKA, which combine dipoles into tiles that are then cross-correlated, the FFTT performs an FFT to cross-correlate all the dipoles giving it considerably greater sensitivity.  For the MWA and SKA, we assume that the effective collecting area of the instrument scales as $\lambda^2$, as expected for a dipole antennae.
We summarize our choice of 21 cm experiment parameters in Table \ref{tab:21cm}.  For each experiment, we impose a cutoff at $k=2\pi/y$, where $y$ is the conformal depth of the redshift slice, to account for the presence of foregrounds, whose removal will reduce the information available on large scales \citep{mcquinn2005}.  We also impose $k_{\rm max}=3h\, {\rm Mpc}^{-1}$ as the non-linear scale and exclude larger $k$ values from the analysis. 

\begin{table}[htdp]
\caption{Low-frequency radio telescopes and their parameters.  For MWA we assume a single redshift slice centered at $z=8$, while for SKA and FFTT we divide the full redshift range into five redshift slices of thickness $\Delta z\approx0.5$.}
\begin{center}
\begin{tabular}{c|cccc|ccc}
Array & $N_a$ & $A_{\rm tot}$ & $ D_{\rm min}$ & $D_{\rm max}$ & $B$& $T_{\rm int}$& $z$\\
          &	&$(10^3\,{\rm m^2})$&$({\rm m})$ & $({\rm km})$ &  (MHz)  &  (hr) \\
\hline
MWA & 500 & 7.0 & 4 & 1.5 & 8 & 4000  &  7.8-8.2\\
SKA & 5000 & 600 & 10 & 5 & 8 & 4000  & 7.8-10.3\\
FFTT & $10^{6}$ & $10^3$ & 1 & 1 & 8 & 4000  & 7.8-10.3\\
\end{tabular}
\end{center}
\label{tab:21cm}
\end{table}%

\begin{figure}[htbp]
\begin{center}
\includegraphics[width=8.6cm, height=8.6cm]{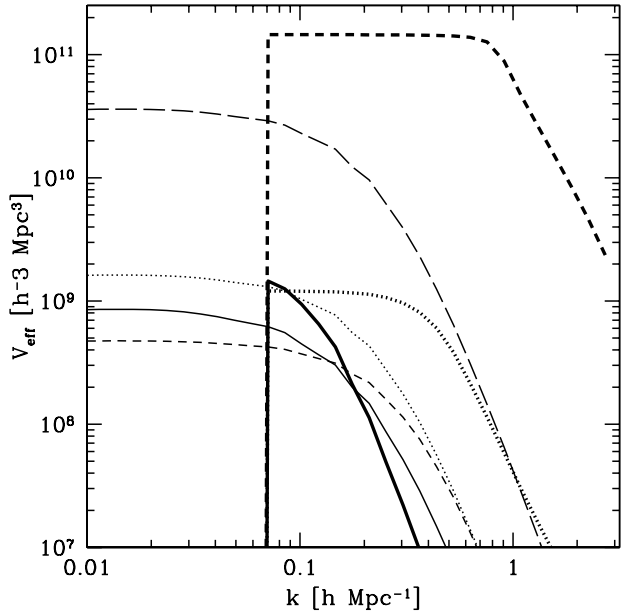}
\caption{Comparison of angle averaged effective volume for galaxy and 21 cm surveys.  SDSS (thin solid curve), G1 (thin dotted curve), G2 (thin short dashed curve), G3 (thin long dashed curve), MWA (thick solid curve), SKA (thick dotted curve), and FFTT (thick dashed curve).  }
\label{fig:veff}
\end{center}
\end{figure}

To give a detailed comparison of these experiments, and so better understand why they result in different parameter constraints, it is useful to define and plot the effective volume observed.  It has previously been pointed out that 21 cm experiments potentially explore a larger volume \citep{LW08}.  The effective volume is defined explicitly for galaxy surveys in Equation \eqref{veff} and we define an equivalent quantity for 21 cm experiments (essentially weighting the survey volume by the signal to noise) by analogy to Equation \eqref{fisher_gal}
\begin{equation}
V_{\rm eff}^{\rm 21cm}(k,\mu)=\frac{P_{T_b}^2}{\sigma_P^2(k,\mu)}V_{\rm survey}.
\end{equation}
The different effective areas for the experiments that we consider are plotted in Figure \ref{fig:veff}.  Looking first at the galaxy surveys, we see that G1 probes more volume on all scales than SDSS leading to improved parameter constraints.  G2 is less sensitive on larger scales, but is able to probe deeper than SDSS leading to extra constraining power on parameters, like $M_\nu$, which affect small scales.  MWA's concentrated antennae distribution gives it good sensitivity to modes $k\sim0.1\iMpc$, but drops in sensitivity dramatically on small scales, where it is noise dominated.  SKA's greater collecting area and more dispersed layout allows it to constrain scales $k\sim1\iMpc$.  The FFTT has a much larger effective volume than any of these experiments and is effectively sample variance limited out to $k\sim 1\iMpc$.  This tremendous volume gives it the power to constrain cosmological parameters at unprecedented levels.  It is instructive to compare it with G3, which probes similarly large volumes, but whose sensitivity drops rapidly on scales $k\gtrsim0.1\iMpc$.  This different behavior makes all the difference when it comes to constraining $M_\nu$.  One unfortunate feature is that foreground removal, which removes the large scales, is likely to prevent a 21 cm experiment observing the imprint of neutrinos on the power spectrum at both large scales ($k\sim0.001 h\iMpc$), where there is little effect, and small scales  ($k\gtrsim0.1 h\iMpc$), where the power spectrum is highly suppressed.  

Why is there this difference in sensitivity on small scales?  For a galaxy survey to have useful sensitivity it is important that there be a large enough density of target galaxies, i.e. we require $P(k)\bar{n}\gtrsim1$.  If the observed galaxies are sparsely distributed within the survey volume, then they only poorly trace small scale features in the dark matter power spectrum.  Consequently, galaxy surveys are at the mercy of the details of galaxy formation and lose sensitivity once pushed to scales far below the typical galaxy spacing.  In contrast, the sensitivity of 21 cm experiments on small scales is a strong function of the distribution of antennae -- a design feature under human control.  If the array lacks a sufficiently high density of long baselines it becomes noise dominated at large $k$ values.  By close filling the array to large baselines, as in FFTT, sensitivity of a 21 cm experiment can be large for even very large $k$ values.  Since hydrogen gas is present on all scales there is no need to worry about an equivalent shot noise component.  While the small scale 21 cm power spectrum might be modified on small scales by astrophysical effects, given the high sensitivity achievable it is worth exploring the use of 21 cm studies further.

\section{Results}
\label{sec:results}

\subsection{Total neutrino mass}
\label{ssec:total}

Table \ref{tab:errors} summarizes the Fisher matrix parameter constraints for galaxy surveys and 21 cm experiments alone and in combination with Planck.  We also list results for a CosmicVar and FFTT.  Our parameter constraints on $M_\nu$ are consistent with the predictions of \citep{lesgourgues2004} for Planck and SDSS. 

As we have shown 21 cm experiments benefit from accessing a larger effective volume than most galaxy surveys. This is mostly a consequence of the well known fact that there is more volume available at higher redshift.  The MWA performs comparatively to large galaxy surveys such as G1 and G2.  With the SKA we begin to exploit the benefit of larger volume seeing the neutrino constraint halve to $\sigma_{M_\nu}=0.16$.  At this level, SKA, by itself, could detect our fidicial neutrino mass $M_\nu=0.3$, although barely at the $2-\sigma$ level.  Adding in Planck to MWA and SKA leads to substantial improvement and allowing a $4-\sigma$ detection of neutrino mass.  The FFTT's high sensitivity allows it to constrain neutrino mass at the level of $\sigma_{M_\nu}=0.01$ by itself.  This is sufficient to detect either normal or inverted mass hierarchies and would open up an era of precision cosmological measurements of neutrino mass.  In general, combining Planck with other 21 cm experiments helps with $M_\nu$ constraints, although this is a very weak effect for FFTT.  Planck is also vital for breaking the degeneracy between bias and $A_S$.

\begin{table*}[htdp]
\caption{Fiducial parameter values and $1-\sigma$ experimental uncertainties for basic 8 parameter model.  Dashes indicate parameters not relevant for that experiment; $\infty$ indicates parameters that are relevant, but not constrained. }
\begin{center}
\begin{tabular}{c|cccccccc|c|cccccc}
\hline
& $\Omega_mh^2$ & $\Omega_bh^2$ & $\Omega_\Lambda$ & $w$ & $n_s$ & $A^2_s$ & $\tau$ & $Y_{He}$ & $M_\nu$ & $b_{SDSS}$ & $T_b(8)$& $T_b(8.5)$& $T_b(9)$ & $T_b(9.5)$ & $T_b(10)$ \\
\hline
Fiducial& 0.15 & 0.023 & 0.7 &-1 & 0.95 & 27 & 0.1 & 0.24 & 0.3 & 2.9 & 26 & 26 & 27 &27 &28\\
\hline
SDSS& 0.154 & 0.0363 & 0.0548 & - & 0.425 & $\infty$  & - & - & 2.54 & $\infty$  & - & - & - & - & -  \\
G1& 0.0294 & 0.00722 & 0.00988 & - & 0.07 & $\infty$ & - & -& 0.431 & $\infty$ & - & - & -  & -  & -  \\
G2& 0.0216 & 0.0058 & 0.00648 & - & 0.0321 & $\infty$  & - & - & 0.371 & $\infty$  & - & - & - & - & -  \\
G3& 0.00623 & 0.00153 & 0.0021 & - & 0.015 & $\infty$ & - & - & 0.091 & $\infty$  & - & - & -  & - & - \\
MWA & 0.0279 & 0.0075 & 0.0139 & - & 0.046 & $\infty$  & - & -  & 0.57  & - & $\infty$ & - & -  & -& -\\
SKA & 0.00043 & 0.00056 & 0.0019 & - & 0.005 & $\infty$ & - & -  & 0.16 & - & $\infty$ & $\infty$ & $\infty$ & $\infty$ & $\infty$  \\
FFTT& 3.8e-05 & 4.77e-05 & 9.18e-05 & - & 0.0002 & $\infty$ & - & - & 0.0089 & - & $\infty$ & $\infty$ & $\infty$ & $\infty$ & $\infty$  \\
\hline
Planck & 0.0045 & 0.00022 & 0.04 & - & 0.0071 & 0.26 & 0.0048 & 0.01 & 0.38  & - & - & - & - & - & -  \\
\hline
+SDSS& 0.00236 & 0.000204 & 0.0216 & - & 0.0068 & 0.25 & 0.00448 & 0.01 & 0.217& 0.197  & - & - & - & -  & - \\
+G1 & 0.000924 & 0.000196 & 0.00749 & - & 0.0066 & 0.243 & 0.00439 & 0.00989  & 0.101  & 0.067 & - & - & - & - & -\\
+G2 & 0.00081 & 0.0002 & 0.0063 & - & 0.0063 & 0.24 & 0.0044 & 0.0092  & 0.10  & 0.12  & - & - & - & - & -\\
+G3& 0.00037 & 0.00016 & 0.0016 & - & 0.0051 & 0.24 & 0.0043 & 0.0081 & 0.046  & 0.029 & - & - & - & -  & -\\
+MWA& 0.0013 & 0.0002 & 0.011 & - & 0.0063 & 0.24 & 0.0044 & 0.0095 & 0.13 & - & 0.74  & - & - & - & -  \\
+SKA& 0.00029 & 0.00014 & 0.0016 & - & 0.003 & 0.23 & 0.0043 & 0.0041 & 0.075 & - & 0.46 & 0.47 & 0.48 & 0.49 & 0.49  \\
+FFTT& 3.72e-05 & 3.58e-05 & 9.13e-05 & - & 0.0002 & 0.23 & 0.0043 & 0.003 & 0.0075 & - & 0.12 & 0.12 & 0.12 & 0.13 & 0.13  \\
\hline
CosmicVar & 0.00148 & 4.12e-05 & 0.014 & - & 0.0024 & 0.12 & 0.0021 & 0.0027 & 0.14 & - & - & - & -  & - & -  \\
+FFTT& 3.28e-05 & 2.26e-05 & 8.86e-05 & - & 0.0002 & 0.11 & 0.0021 & 0.00096& 0.0059 & - & 0.06 & 0.06 & 0.07 & 0.07 & 0.07  \\
\end{tabular}
\end{center}
\label{tab:errors}
\end{table*}%

\begin{figure}
\centering
\includegraphics[width=8.6cm, height=8.6cm]{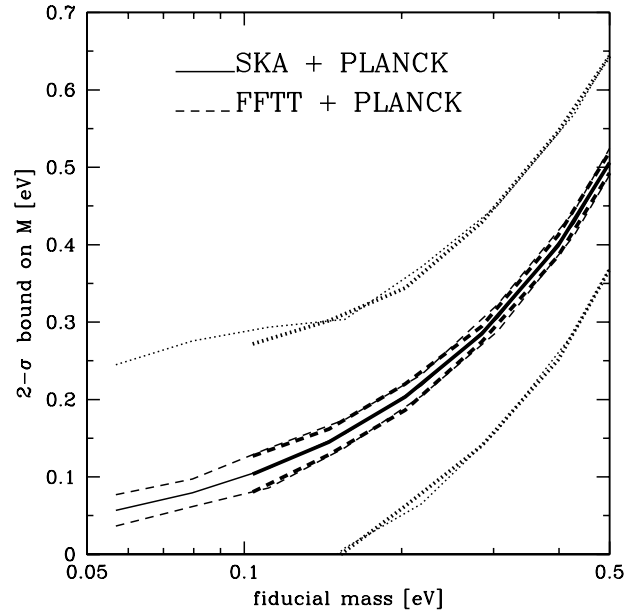}
\caption{\label{fig:massplot}  2$\sigma$ constraints on $M_\nu$ (solid curve) from SKA+PLANCK (dotted curve) and FFTT+PLANCK (dashed curve) for the normal (thin curves) and inverted (thick curves) hierarchies. }
\end{figure}

Figure \ref{fig:massplot} illustrates the effect of changing $M_\nu$ on the parameter constraints.  Since only SKA and FFTT are capable of placing useful constraints on $M_\nu$ we restrict to plotting curves for these experiments.  In general, the mass constraints are relaxed as we move to smaller $M_\nu$, where the impact of neutrino mass is lessened, but the constraint does not depend upon the hierarchy.

Based upon these results, SKA+Planck should be capable of detecting neutrino mass at the $2-\sigma$ mass provided that $M_\nu\gtrsim0.15\eV$.  FFTT+Planck is capable of detecting neutrino mass for all masses above the minimum mass in both normal and inverted hierarchies.  Further, FFTT+Planck has sufficient sensitivity to be sensitive to the hierarchy itself.  From Figure \ref{fig:massplot}, we see that it could measure masses in the range $M_\nu<0.11 \eV$, below the minimum mass allowed for the inverted hierarchy and so confirm the existence of a normal hierarchy with $M_\nu$ in this range. 

Table \ref{tab:werrors} summarizes the Fisher matrix parameter constraints for galaxy surveys and 21 cm experiments alone and in combination with Planck when the equation of state $w$ is allowed to vary.  This significantly weakens the constraints from galaxy surveys, but has little effect on the neutrino constraints for 21 cm experiments.  Since 21 cm experiments are generally well within the matter dominated regime they are only sensitive to the dark energy through the angular diameter distance.  This is essentially the same measurement as made by the CMB. Once Planck is added to galaxy or 21 cm surveys very similar $M_\nu$ constraints as in the case of $\Lambda$CDM are obtained.  This extra data set is very powerful at breaking the degeneracy.

\begin{table*}[htdp]
\caption{Fiducial parameter values and $1-\sigma$ experimental uncertainties for basic 8 parameter model+variation in $w$.  Dashes indicate parameters not relevant for that experiment; $\infty$ indicates parameters that are relevant, but not constrained. }
\begin{center}
\begin{tabular}{c|cccccccc|c|cccccc}
\hline
& $\Omega_mh^2$ & $\Omega_bh^2$ & $\Omega_\Lambda$ & $w$ & $n_s$ & $A^2_s$ & $\tau$ & $Y_{He}$  & $M_\nu$ & $b_{SDSS}$ & $T_b(8)$& $T_b(8.5)$& $T_b(9)$  & $T_b(9.5)$& $T_b(10)$\\
\hline
Fiducial& 0.147 & 0.023 & 0.7 & -1 & 0.95 & 26.6 & 0.1 & 0.24 & 0.3 & 2.3 & 26 & 26 & 27 &27 &28 \\
\hline
SDSS& 0.456 & 0.083 & 0.117 & 1.21 & 0.503 & $\infty$ & - & - & 6.16 & $\infty$ & - & - & -& - & -   \\
G1& 0.119 & 0.0207 & 0.0358 & 0.574 & 0.174 & $\infty$ & - & -& 1.28 & $\infty$ & - & - & - & - & -  \\
G2 & 0.0354 & 0.00593 & 0.295 & 1.22 & 0.0482 & $\infty$ & - & - & 1.01 & $\infty$ & - & - & - & - & - \\
G3& 0.0252 & 0.00438 & 0.0076 & 0.122 & 0.037 & $\infty$& - & - & 0.272 & $\infty$ & - & - & -  & - & - \\
MWA & 0.0317 & 0.00761 & 0.972 & 3.13 & 0.0487 & $\infty$ & - & - & 0.749 & - & $\infty$ & - & - & - & - \\
SKA& 0.00191 & 0.00056 & 0.234 & 0.747 & 0.0054 & $\infty$ & - & - & 0.175 & -  & $\infty$ & $\infty$ & $\infty$ & $\infty$ & $\infty$  \\
FFTT& 4.94e-05 & 4.77e-05 & 0.0045 & 0.014 & 0.0002 & $\infty$ & - & - & 0.009 & - & $\infty$ & $\infty$ & $\infty$ & $\infty$ & $\infty$  \\
\hline
Planck& 0.0045 & 0.00024 & 0.068 & 0.18 & 0.0074 & 0.26 & 0.0048 & 0.011 & 0.38 & - & - & - & -  \\
\hline
+SDSS& 0.0033 & 0.00024 & 0.023 & 0.11 & 0.0074 & 0.254 & 0.0046 & 0.0103 & 0.272& 0.197  & - & - & - & - & -\\
+G1 & 0.0016 & 0.00021 & 0.013 & 0.081 & 0.0068 & 0.245 & 0.0044 & 0.01& 0.136 & 0.067 & - & - & - & - & -  \\

+G2& 0.00089 & 0.00022 & 0.037 & 0.149 & 0.0067 & 0.243 & 0.0044 & 0.0099& 0.104 & 0.16 & - & - & - & - & -   \\
+G3& 0.00051 & 0.00016 & 0.003 & 0.021 & 0.0051 & 0.24 & 0.0043 & 0.0081& 0.052 & 0.029 & - & - & - & -   & - \\
+MWA& 0.00146 & 0.00021 & 0.053 & 0.17 & 0.0066 & 0.242 & 0.0044 & 0.0101& 0.144 &- &   1.45 & - & - & - & -  \\
+SKA& 0.00029 & 0.00014 & 0.020 & 0.065 & 0.003 & 0.236 & 0.0043 & 0.0044 & 0.080 &- & 0.71 & 0.72 & 0.74 & 0.75 & 0.76  \\
+FFTT& 4.23e-05 & 3.97e-05 & 0.004 & 0.011 & 0.0002 & 0.23 & 0.0043 & 0.0030 & 0.0075 & - & 0.14 & 0.15 & 0.15 & 0.15 & 0.16 \\
\hline
CosmicVar& 0.00244 & 4.16e-05 & 0.030 & 0.033 & 0.0024 & 0.124 & 0.0023 & 0.0028& 0.222 & - & - & - & - & - & -  \\
+FFTT& 3.3e-05 & 2.58e-05 & 0.0024 & 0.0076 & 0.0002 & 0.111 & 0.0021 & 0.0011 & 0.0068 & - & 0.09 & 0.09 & 0.1 & 0.1 & 0.1  \\
\end{tabular}
\end{center}
\label{tab:werrors}
\end{table*}%

\begin{figure}[htbp]
\begin{center}
\includegraphics[width=8.6cm, height=8.6cm]{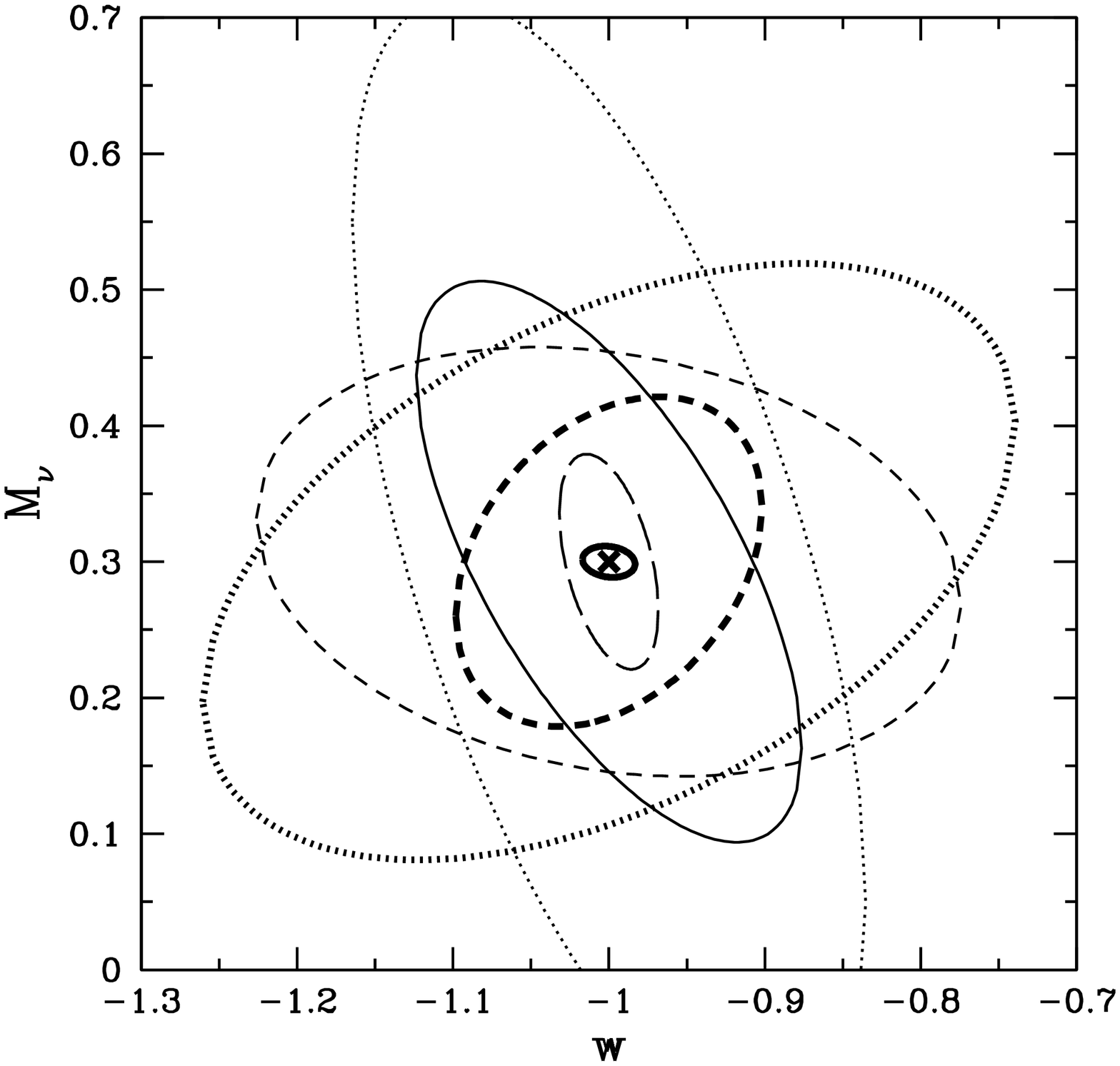}
\caption{ 68-\% confidence ellipses in the $M_\nu-w$ plane from PLANCK in combination with SDSS (thin dotted), G2 (thin short dashed), G1 (thin solid), G3 (thin long dashed), MWA (thick dotted), SKA (thick dashed), and FFTT (thick solid).  }
\label{fig:eplot_mnu}
\end{center}
\end{figure}
We summarize these result in Figure \ref{fig:eplot_mnu}, where we plot $68-\%$ confidence intervals in the $w-M_\nu$ plane.  The contours for galaxy surveys show a clear diagonal alignment indicating significant degeneracy between $w$ and $M_\nu$.  Hence better measurements of either quantity translate into better measurements of the other.  In contrast, the contours for 21 cm experiments are less closely aligned with the axes indicating little degeneracy.  This is as expected since the 21 cm experiments are focused at high redshifts where dark energy has little effect.  This also explains why 21 cm experiments are generally quite poor at constraining $w$.  Note that since these observations take place at very different redshifts the axes of the error ellipse are not aligned implying that combining all data sets would further reduce the maximum constraint achievable.

Finally, it is worth noting that constraints on $Y_{\rm He}$ may themselves be interesting from the point of view of constraining neutrino physics.  In this analysis, as in most, we have treated $Y_{\rm He}$ as a free parameter to be varied independently.  In the framework of cosmology, however, once $\Omega_b h^2$ is known predictions for $Y_{\rm He}$ can be made using the machinery of Big Bang Nucleosynthesis (BBN).  A discrepancy between the prediction and observation would point to an early modification of general relativity or, more interesting for neutrinos, the presence of extra relativistic degrees of freedom $N_{\nu,eff}$ during BBN, such as a sterile neutrino.  The power of this additional constraint was pointed out by \citep{hamann2008} and here we simply comment on the applicability of 21 cm observations.  Current constraints on $Y_{\rm He}$ come from observations of Helium lines in metal poor HII regions.  By extrapolating the observed abundances back to zero metallicity a primordial abundance can be estimated.  These observations place bounds $Y_{\rm He}=0.2477\pm0.0029$ \citep{peimbert2007}, but there is considerable systematic uncertainty relating to, amongst other problems, the unknown temperature profile of the HII region.  Given these systematic effects it is unclear that these constraints can be improved much beyond an additional factor of two further.  Additional constraints from more reliable sources are clearly important.  Table \ref{tab:errors} shows that Planck only weakly constrains \yhe, since \yhe modifies the power spectrum primarily on small scales where Silk damping reduces the amplitude of the signal.  Adding additional information on small scales through 21 cm observations helps break this degeneracy leading to factor of 2 (3) improvements for SKA (FFTT).  The constraint from Planck + FFTT of 0.0028 is comparable with current observational bounds.  Note that although, through the baryon component, \yhe affects the matter power spectrum on small scales the effect is very small.  We find that by themselves SKA (FFTT) can place constraints on \yhe of 0.17 (0.014).  This extra information offers only a small improvement over that contained in the CMB.

%%%%%%%%%%%%%%%%%%%%%%%%%%%%%%%%%%%%%%%%%%%
\subsection{Neutrino mass splitting}
\label{ssec:splitting}

We have seen in the previous subsection that SKA, G3, and FFTT all possess the sensitivity needed to provide a detection of neutrino mass at high sensitivity (at least for some range of $M_\nu$).  It is interesting to ask if any of these experiments can go further and begin to constrain the masses of individual neutrino mass states.  As mentioned in \S\ref{sec:nugen} the power spectrum is primarily sensitive to the total neutrino energy density, but is modified at the $0.2\%$ level by mass differences between the three neutrino states.  Precise power spectrum measurements could, in principle, make this measurement.  This would conclusively identify neutrinos as acting in a cosmological context.

\begin{figure}[htbp]
\begin{center}
\includegraphics[width=8.6cm, height=8.6cm]{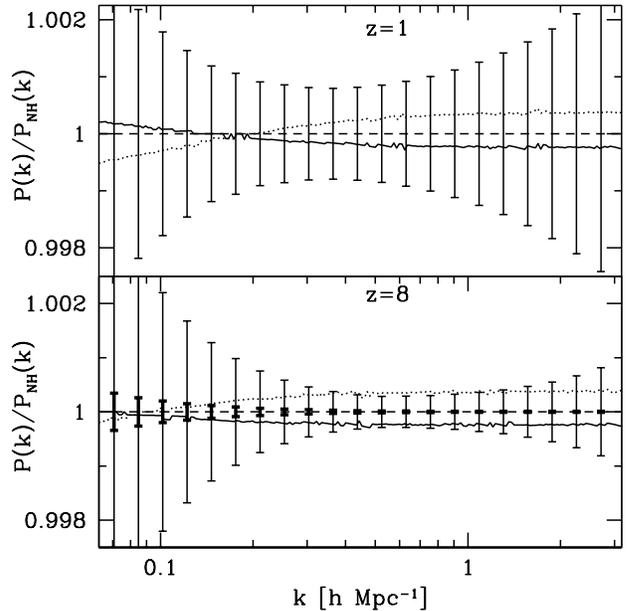}
\caption{{\em Top panel: }1-$\sigma$ fractional errors on the power spectrum at $z=1$ for G3.  {\em Bottom panel: }1-$\sigma$ fractional errors on the power spectrum at $z=8$ for SKA (thin black points), and FFTT (thick black points).  In each panel, we plot the ratio of $P(k)$ for three degenerate masses (dotted curve) or inverted hierarchy (solid curve) to $P(k)$ for the normal hierarchy neutrinos (as in Figure \ref{fig:ratio}) and $M_\nu=0.3\eV$. Only the FFTT has the sensitivity to detect the mass splittings at a significant level. Nonlinearities are expected to affect the $z=1$ power spectrum on scales smaller than $k\gtrsim0.2h\iMpc$ and the $z=8$ power spectrum on scales $k\gtrsim3 h\iMpc$.}
\label{fig:hratio_errors}
\end{center}
\end{figure}

We begin our analysis by plotting the $1-\sigma$ power spectrum errors for G3, SKA, and FFTT, which are shown in Figure \ref{fig:hratio_errors}.  In each case, we consider the errors on a fiducial model with $M_\nu=0.3$ distributed in the normal hierarchy (dashed line).  For comparison, we plot curves for the power spectrum in the case of the inverted hierarchy (dotted curve) and for three neutrinos with degenerate masses (solid curve).  These curves give an indication of how precise power spectrum measurements must be to have a hope of distinguishing individual neutrino masses.  In the top panel of Figure \ref{fig:hratio_errors}, we plot errors for G3 on the power spectrum at $z=1$.  It is clear that while G3 measures the power spectrum at the $0.1\%$ level it is still not precise enough to distinguish the different mass splittings.  This is consistent with the findings of \citet{lesgourgues2004}.  Note that non-linearities in galaxy bias and the underlying density field would interfere 
with measurements of the power spectrum on scales $k>0.2 h\iMpc$ anyway, further restricting the ability of a low redshift galaxy survey to constrain neutrino mass splittings.

In the bottom panel of Figure \ref{fig:hratio_errors}, we consider power spectrum errors from SKA and FFTT at $z=8$.  First note that the differences between the three different models has different scale dependence than at $z=1$.  Unfortunately, this does not greatly increase the chances of detecting the mass splitting.  Next, since the non-linear scale at $z=8$ is $k\gtrsim3h \iMpc$ the whole of the plotted region is potentially accessible to experiment without having to worry about complications of a non-linear density field.  We will discuss complications arising from small scale modifications due to inhomogeneous radiation fields to the 21 cm power spectrum later.  SKA has almost the sensitivity needed to differentiate between different mass splittings at the $1-\sigma$ level on scales $0.4 \le k \le 1 h\iMpc$.  The FFTT does visibly better, having error bars smaller than the difference in the three power spectra on scales  $k \ge 0.2 h\iMpc$.  This motivates the next part of our analysis, where we analyze the Fisher matrix constraints on individual neutrino masses for these experiments, since degeneracies between different parameters affect the detectability of neutrino mass.

\begin{table}[htdp]
\caption{Neutrino mass $1-\sigma$ uncertainties.  These uncertainties are calculated using the full 7 parameter plus three independent neutrino masses Fisher matrix and then projected onto the reduced parameter set listed.}
\begin{center}
\begin{tabular}{c|ccc|cc|c}
& $m_1$ & $m_2$ &$m_3$ & $m_1+m_2$  &$m_3$ &$M_\nu$\\
\hline
Fiducial & 0.081 & 0.09 & 0.13 & 0.17& 0.13 &0.3 \\
\hline
Planck & 0.922 & 0.950 & 0.552 & 0.633 & 0.550 & 0.44 \\
+G3 & 0.654 & 0.78 & 0.295 & 0.296 & 0.265 & 0.046\\
+SKA & 0.645 & 0.631 & 0.405 & 0.41 & 0.405 & 0.077\\
+FFTT & 0.037 & 0.03 & 0.03 & 0.03 & 0.029 & 0.0076\\
\hline
Cosmic Variance & 0.285 & 0.282 & 0.209 & 0.235 & 0.209 & 0.157 \\
+G3 & 0.247 & 0.272 & 0.139 & 0.15 & 0.137 & 0.027\\
+SKA& 0.265 & 0.264 & 0.194 & 0.192 & 0.194 & 0.037\\
+FFTT & 0.036 & 0.03 & 0.03 & 0.029 & 0.028 & 0.006\\
\end{tabular}
\end{center}
\label{tab:m3errors}
\end{table}%

We now examine in detail the ability of our fiducial experiments to constrain individual neutrino masses.  We take as our fiducial model three neutrino mass states distributed according to the normal hierarchy with $M_\nu=0.3\eV$.  This fiducial mass is sufficiently large that a measurement of $M_\nu$ alone would not allow either hierarchy to be ruled out.  Only if individual neutrino masses could be measured could the hierarchy be established.  

While CAMB has sufficient precision to calculate the effect on the power spectrum of individual neutrino masses, the resulting differences in the power spectrum are only slightly larger than numerical uncertainties in the code (just visible as numerical noise in Figure \ref{fig:ratio}).  We address this by running CAMB at its highest accuracy settings and visually inspecting the power spectrum derivatives to ensure that there are no distinctive glitches.  Large numerical noise could easily break the degeneracy between different neutrino masses yielding erroneously small error bars.  As a numerical check, by projecting from errors on ($m_1$, $m_2$, $m_3$) onto $M_\nu$ we see that these results are consistent with the results of the previous section.  

Parameter constraints are shown in Table \ref{tab:m3errors} for individual neutrino masses.  From the fisher matrix for three individual neutrino masses, we project onto the \{$m_1+m_2$, $m_3$\} and \{$M_\nu$\} parameter spaces.  Since the mass splitting between $m_1$ and $m_2$ is $\lesssim0.01\eV$ there exists a strong degeneracy between these two masses for most experiments.  In contrast the sum of these two nearly degenerate masses is much better constrained.  By considering the sum, we get a better handle on whether the experiments could distinguish between normal and inverted hierarchy by isolating whether one or two neutrino mass states lie above the largest mass splitting. 

Neither G3 or SKA is able to place significant constraints on the individual neutrino masses.   FFTT is able to distinguish from zero all three mass states at the $>3-\sigma$ level.  We conclude that an FFTT type 21 cm experiment in conjunction with at least Planck level CMB data might someday make a concrete detection of individual neutrino masses in the cosmological context.  

\begin{figure}[htbp]
\begin{center}
\includegraphics[width=8.6cm, height=8.6cm]{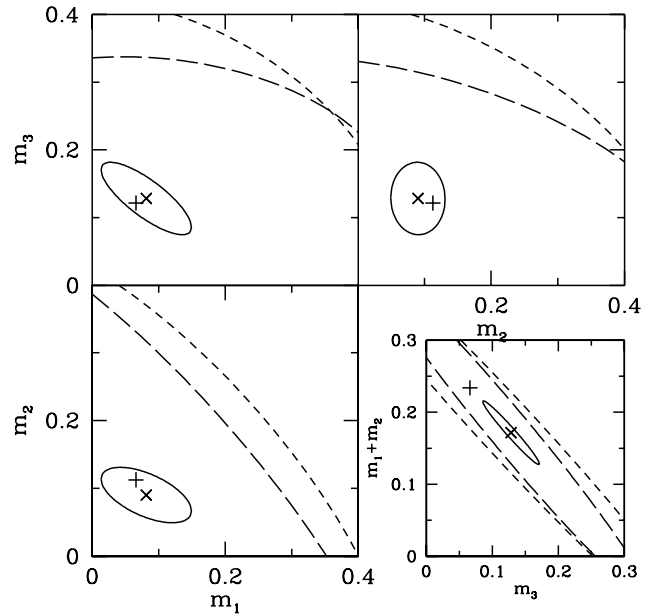}
\caption{68\% confidence contours on individual neutrino mass states $m_1$, $m_2$, and $m_3$.  We show contours for CosmicVar in combination with G3 (long dashed curve), SKA (short dashed curve), and FFTT (solid curve). Fiducial neutrino masses are indicated for the normal hierarchy (diagonal crosses), also shown are the values for the inverted hierarchy (vertical crosses) for $M_\nu=0.3\eV$.   CosmicVar+FFTT is able to make a weak detection of individual neutrino masses. Insert in the bottom right hand corner shows the error ellipse in the $m_1+m_2$-$m_3$ parameter space and illustrates the strong degeneracy hampers placing tight constraints on individual masses. }
\label{fig:eplot_m3}
\end{center}
\end{figure}

For comparison with these results, the inverted hierarchy with $M_\nu=0.3\eV$ satisfies ($m_1$, $m_2$, $m_3$)=(0.066,0.113,0.122) eV differing from the normal hierarchy by ($\Delta m_1$, $\Delta m_2$, $\Delta m_3$)=(0.015,0.023,0.008) eV.  These differences are too small for any of the experiments to distinguish between the two hierarchies (see Figure \ref{fig:eplot_m3}).  Thus although individual neutrino masses may be measured, the precision required to distinguish normal and inverted hierarchies directly at the level of individual masses is difficult to achieve.  It would be possible in principle by increasing the collecting area or integration time of the FFTT, although the design we consider is already quite futuristic.  

\section{Conclusions}
\label{sec:conc}

For the foreseeable future cosmology is likely to provide insights into neutrino masses, constraining the underlying particle physics.  In this paper, we have explored the possibility of using low-frequency radio observations of the 21 cm line of hydrogen during the epoch of reionization to constrain neutrino masses and compared with future galaxy surveys.  In general, we find that the outlook for 21 cm experiments is promising since these experiments are capable of probing small scale modifications of the matter power spectrum with higher sensitivity than galaxy surveys.  This is in part a consequence of going to higher
redshifts where non-linear modifications to the matter power spectrum are expected to be smaller.  It is also a consequence of the sensitivity of 21 cm experiments being dominated on small scales by the distribution of antennae.  If the radio array is constructed with a large enough density of long base lines it can be made sensitive to large $k$-modes.  In contrast, galaxy surveys are reliant upon there being a large enough density of target galaxies so that power spectrum measurements are not dominated by shot noise at large $k$ values.

We find that 21 cm experiments such as SKA should be competitive with galaxy surveys for constraining the total neutrino mass.  In combination with Planck, SKA should be able to measure the effects of neutrino mass at the $2-\sigma$ level if $M_\nu\gtrsim0.15$.  These constraints are largely independent of uncertainties in the dark energy, which affects lower redshift probes of the matter power spectrum.
In the further future, an instrument with a square kilometer of collecting area able to cross-correlate all its antennae, such as FFTT, could measure $M_\nu$ at the better than $2-\sigma$ for all values of $M_\nu$ currently allowed by the observed mass splittings.  This is exciting as it raises the possibility of using cosmology to distinguish between the normal and inverted neutrino mass hierarchies.  Previous work \citep{lesgourgues2004} had shown that even an all-sky galaxy survey out to $z=1$ is unlikely to be able to measure individual neutrino masses.  We obtain a similar conclusion for a 21 cm experiment such as SKA, but note that a successor instrument, such as FFTT, might indeed be able to distinguish individual masses.  Our calculations show that FFTT could make $\gtrsim2-\sigma$ detections of each of three neutrino mass states.  This would be valuable for making a measurement of neutrino masses, if laboratory experiments have not already done so, or confirming that we are indeed seeing the effect of neutrinos in the cosmological context.

Whether 21 cm experiments live up to the promise shown by this work depends on a number of factors.  First of all is the very real problem of astrophysical foregrounds \citep{dimatteo2002,ohmack2003}, which must be overcome before 21 cm fluctuations can be measured and used for cosmology.  The outlook for foreground removal is positive \citep{zfh2004freq,santos2005} and will be better understood once the first generation of instruments, e.g. LOFAR, MWA, 21CMA, are completed and collect data.  Next is the possibility that modifications to the 21 cm power spectrum from reionization or from astrophysical radiation fields degrade the accuracy with which the underlying density field can be measured.  We have taken the optimistic point of view that this will not affect parameter constraints in order to get a sense of how well one might hope to do. In light of recent work by \citet{mao2008}, it is likely that these effects will be an issue, but not a show stopper.  \citet{mao2008} find that needing to fit for unknown parameters describing reionization degrades cosmological parameter constraints by between $10-50\%$.  This would not prevent the detection of individual neutrino masses that we have described here, although clearly it would reduce their significance.  In summary, it seems possible that future 21 cm experiments will provide highly complementary constraints on neutrino masses and could push beyond those obtained from galaxy surveys.  There are clearly challenges ahead for this technique, but if overcome, the future of neutrino cosmology may be bright.

\acknowledgments
JRP is supported by NASA through Hubble Fellowship grant HST-HF-01211.01-A
awarded by the Space Telescope Science Institute, which is operated by the
Association of Universities for Research in Astronomy, Inc., for NASA,
under contract NAS 5-26555.
  Extensive use of CAMB (http://camb.info) made possible the power spectrum calculations in this paper.
EP is an NSF--ADVANCE fellow (AST--0649899)
also supported by NASA grant NNX07AH59G, JPL--Planck subcontract  no. 1290790 and JPL SURP award no. 1314616. She would also like to thank Caltech for hospitality during the preparation of this work.
Finally, the authors would like to thank the hospitality of the Aspen Center for Physics where this work was initiated.

%%%%%%%%%%%%%%%%%%%%%%%%%%%%%%%%%%%%%%%%%%%%%%%%%%%%%%%%%%%%%%%%%%%%%%%%%%%%%%%%
% \bibliography{neutrinobib,CARbiblio}
%%%%%%%%%%%%%%%%%%%%%%%%%%%%%%%%%%%%%%%%%%%%%%%%%%%%%%%%%%%%%%%%%%%%%%%%%%%%%%%%

%%%%%%%%%%%%%%%%%%%%%%%%%%%%%%%%%%%%%%%%%%%%%%%%%%%%%%%%%%%%%%%%%%%%%%%%%%%%%%%%
 
 \end{document}